# Reduction of wood hygroscopicity and associated dimensional response by repeated humidity cycles

Luis García Esteban*, Joseph Gril**, Paloma de Palacios de Palacios*, Antonio

Guindeo Casasús*

*Cátedra de Tecnología de la Madera. Departamento de Ingeniería Forestal

Escuela Técnica Superior de Ingenieros de Montes. Universidad Politécnica de Madrid. Spain.

Tel. 91.336.71.21, Fax. 91.336.71.26, e-mail: luis.garcia@upm.es

**Laboratoire de Mécanique et Génie Civil, Université Montpellier 2, France

*Running title:*

**Reduction of wood hygroscopicity**

**Abstract.** The reduction of wood response to variations of ambient humidity, described as kind of ageing, has been studied from both points of view of moisture uptake and dimensional changes. Clearwood specimens from 3 gymnosperms (*Pinus sylvestris* L., *Pinus pinaster* Ait., *Pinus insignis* Dougl.) and 4 angiosperms (*Populus spp.*, *Quercus pyrenaica* Willd., *Entandrophragma cylindricum* Sprague, *Chlorophora excelsa* Benth. & Hook f.) were subjected to five wet-dry cycles and their moisture content and dimensional changes in radial and tangential direction at three increasing levels of relative humidity (57.6, 84.2 and 90.2%) were measured before and after the treatment. For a given increase of relative humidity, a coefficient of hygroscopic ageing was defined as the relative decrease of moisture uptake, and the coefficient of dimensional ageing as the difference of swelling strain before and after ageing. All species exhibited a clear ageing effect, much stronger in the wet range than in the dry range. Calculation of swelling coefficients suggested that in some cases the relation between swelling strain and moisture content is no longer linear in aged wood.

# "Réduction de l'hygroscopicité du bois et des variations dimensionnelles associées au moyen de cycles d'humidité répétés"

**Résumé.** La réduction de la réponse du bois aux variations d'humidité ambiante, décrite comme un forme de vieillissement, a été étudiée du double point de vue de la reprise d'humidité et des variations dimensionnelles. Des échantillons de bois sans défaut de 3 gymnospermes (*Pinus sylvestris* L., *Pinus pinaster* Ait., *Pinus insignis* Dougl.) et 4 angiospermes (*Populus spp.*, *Quercus pyrenaica* Willd., *Entandrophragma cylindricum* Sprague, *Chlorophora excelsa* Benth. & Hook f.) ont été soumis à cinq cycles humide-sec à 50°C. Leur taux d'humidité et leurs variations dimensionnelles dans les directions radiale et tangentielle à trois niveaux croissants d'humidité relative (57.6, 84.2 et 90.2%) ont été mesurés avant et après traitement. Pour un accroissement donné d'humidité relative, un coefficient de vieillissement hygroscopique a été défini par la décroissance relative de reprise d'humidité, et un coefficient de vieillissement dimensionnel par la différence de gonflement avant et après traitement. Toutes les espèces exprimaient clairement un effet de vieillissement, plus marqué dans la plage humide que dans la plage sèche. Les calculs de coefficients de gonflement ont suggéré que dans certains cas la relation entre déformation de gonflement et taux d'humidité n'est plus linéaire dans le bois ayant subit un vieillissement en sorption.

**Mots-clés : Bois / sorption / hygroscopicité / gonflement / vieillissement**

# 1. Introduction

If dry wood, whether in an anhydrous state or not, is exposed to hygrothermal conditions corresponding to an equilibrium moisture content greater than that initially present, the wood will start to admit water molecules from the surrounding environment. The vast number of dynamic hygroscopic states of equilibrium through which wood passes means that if the temperature is kept constant, the resulting sorption isotherm exhibits a sigmoid shape as shown in Fig. 1. If the relative humidity is increasing, the trajectory formed is termed sorption isotherm [2] [4] [5]. On the other hand, if it is decreasing, it is termed desorption isotherm. The two sigmoids do not coincide, so that humidity cycles produce hysteresis loops [14] [23] [24] (Fig. 2); the quotient between the sorption and desorption moisture contents, for a single hygrothermal state, has been given the name of hysteresis coefficient [7].

Practice in the field has led experienced carpenters and woodworkers in general to state that the longer the wood is in place on site, the better it is. This is demonstrated by the fact that pieces of wood obtained from structural elements of old buildings have lower contraction coefficients. Physically this phenomenon does not only correspond to an increase in the hysteresis coefficient, getting close to 1, but also to a combined mechanism which has been termed hygroscopic fatigue of wood by some authors [8] [9] [10]. However, the concept of "fatigue" is commonly used in materials science to designate a negative evolution of material properties, associated with the development of some microscopic damage. For that reason, we will prefer the term of "ageing", which is more neutral and can also suggest an improvement caused by irreversible changes in the microstructure, essentially of chemical origin. In essence, the phenomenon of hygroscopic ageing will be defined as the loss of hygroscopic response on the part of the wood after a process of extenuation of the cell wall polarity [17] [19]. It most likely corresponds to a partial saturation among the polar groups of the cell wall, leaving them unavailable for water vapour fixing. Transferred to a sorption or desorption curve, this results into lower slopes of the sigmoid (Fig. 3). Similar observations could be made if the dimension of specimens had been considered, instead of their weight, so that this phenomenon can be seen at two levels, hygroscopic and dimensional. *Hygroscopic and dimensional ageing of wood* will be defined here as the loss of response of wood in relation to hygrothermal changes. This type of ageing is probably irreversible, which makes it in principle different from the reversible "physical ageing" [22], suspected to occur in wood also as a consequence of humidity changes [12].

To evaluate this phenomenon, we based the study on the sorption isotherms corresponding to 25° C for each of the woods used, considering the occurrence of ageing to be shown when the slope of the isotherms decreased in comparison with non-aged wood. This implies that the aged wood will respond with a smaller difference in moisture content (or dimensional change) for a given variation of relative humidity. This is illustrated in Fig. 4 where due to the hygroscopic ageing the wood responds with a lower amplitude to a given step of relative humidity, resulting in a lower slope of the isotherm. A similar approach will be taken for characterizing the dimensional ageing.

## 2. Materials and methods

**2.1 Material**

The species used were: 3 gymnosperms, *Pinus sylvestris* L. (Ps), *Pinus pinaster* Ait. (Pp), *Pinus insignis* Dougl. (Pi) ; 4 angiosperms, *Populus spp.* (pop) ; *Quercus pyrenaica* Willd. (Qp); *Entandrophragma cylindricum* Sprague (Ec); *Chlorophora excelsa* Benth. & Hook f. (Ce). From a total of ten trunks of each species, with an average length of 2.25 m, and an average diameter from 0.35 m to 0.65 m, ten slices of 50 mm thick were obtained from each. A first cut was made on the sections or slices with a band saw, dividing them into two halves following a radial direction, with this being the surface area to be used for obtaining radial test pieces. A second cut was then made on one of the halves, parallel to the previous cut. This was a tangential cut, to be used for obtaining tangential test pieces. The surface areas obtained in this way were laminated while green following the radial and tangential cuts to a thickness of 1 mm using a laboratory roller. Finally, the veneers thus obtained were shorn to attain the final measurements of the test pieces (50x10x1 mm). The test pieces chosen were all from heartwood (Fig. 5).

For all the woods employed in this study, a prerequisite was established that they should all come from a specimen of green lumber in order to guarantee that the recently cut wood had not undergone any natural ageing cycle which may have distorted the study of the physical phenomenon. The justification for drying the wood naturally is based on a number of factors [9] [15] [16] [18]: to avoid checking defects; to avoid warping or internal stresses produced by a poor choice of kiln-drying schedule or through its defective conduction; to eliminate the drying parameter, thereby preventing it from distorting the quantification of the phenomenon.

Moreover, lowering the moisture content of the wood below the fibre saturation point corresponds to the fact that the phenomenon must be demonstrated below this point and above 12 % so that the results obtained can be transferred to the normal values in use.

**2.2. Equipment**

To evaluate the process of hygroscopic ageing, thermostatic baths with water were used, with a regulation of $\pm$ 0.1ºC, equipped with an agitator allowing the temperature to be homogenised throughout the recipient. The flasks on which the test pieces were placed had a Petri dish 3 cm from the free surface area of the saturated solution. To facilitate the conditioning of the baths, the entire process was carried out inside a hygrothermal chamber, as this allowed correct atmospheric homogeneity around the test pieces. It is recommended that the temperature fluctuations be very slight ($\leq \pm 0.2$ºC). To diminish the times required for equilibrium, the process was carried out in a low vacuum (130 mb). To determine the points of hygroscopic equilibrium, a balance with a precision of 0.0001 g was used.

The dimensional ageing was evaluated using a micrometer with a precision of 0.01 mm. To avoid unwanted bowing and warping in some of the test pieces, in particular those originating from very nervous species, the samples were placed between two glass slides 4 mm thick, 70 mm long and 8 mm wide, like a sandwich, and all of this was held together by two clips at the ends so that the test piece extended beyond the ends of the slides. The micrometer measurement was always made on the same point by use of a counterpoint on the opposite side of the measurement being made.

**2.3. Construction of the sorption isotherm**

In constructing the isotherms, use was made of the recommendations of the COST'90 Working Group [21]. To determine the sorption isotherm, the test pieces must be dried to an anhydrous state. This is achieved by placing the test pieces in a desiccator containing $P_2O_5$ for one month at a temperature of 25 ºC. After the anhydrous state is reached, ten anhydrous test pieces, each originating from a different tree, are placed in each desiccator. Hygroscopic equilibrium is reached when the difference between two consecutive weightings (about 45 days) is not significant (0.0005 g). However, before starting to plot the sorption curve, a preliminary calibration should be carried out following the recommendations of the COST'90 Working Group.

Of the salts suggested by the Working Group to achieve the equilibrium points following the isotherm of 25 ºC, those used for the study of ageing were NaBr; KCl and $BaCl_2$, giving a relative air humidity of 57.6 ± 0.4 %, 84.2 ± 0.3 % and 90.19 ± 0.02 % respectively. Once the points of equilibrium are reached, the moisture contents are expressed in grams of water per 100 grams of dry material.

**2.4. Ageing procedure**

Once the sorption isotherm has been constructed, the test pieces are subjected to an accelerated ageing process designed to produce a loss of hygroscopic response in the wood (forced ageing). To achieve this, five alternate ageing cycles were carried out using 50 ºC isotherm and 90 % relative humidity and desiccation to anhydrous weight over a period of three days for each one.

A low ageing temperature was used to avoid modifications in the hygroscopicity of the wood, as high temperatures cause substantial decreases in the percentage of the hemicelluloses [3] [6] [11] and a decrease in the degree of crystallisation of the cellulose [1].

After the five consecutive cycles have been completed, the isotherms are constructed again following the same saturated salts procedure. The number of cycles chosen is in relation to the follow-up performed on the chosen woods during their ageing period. This follow-up showed that all of the woods chosen had experienced a reduction in weight loss and measurements in respect to variations in humidity and temperature at the end of the fifth cycle. At the end of this cycle, the moisture values and lengths in the wood were very similar to those of the fourth cycle.

## 2.5. Quantification of moisture content variations and swelling strains.

The sigmoid-shaped graph which defines the equilibrium moisture content points of wood under different conditions of relative air humidity for a constant temperature (sorption isotherm) suggests to quantify the variation of the moisture content between two points of hygroscopic equilibrium through the use of tangents, or slope of the sorption isotherm (see Fig. 4):

$$a_1^n = \frac{h_2^n - h_1^n}{H_2 - H_1} \quad ; \quad a_2^n = \frac{h_3^n - h_2^n}{H_3 - H_2}$$

$$a_1^f = \frac{h_2^f - h_1^f}{H_2 - H_1} \quad ; \quad a_2^f = \frac{h_3^f - h_2^f}{H_3 - H_2}$$

(1)

where $H_1 = 57.6\%$; $H_2 = 84.2\%$; $H_3 = 90.2\%$ are the three levels of relative humidity applied to the wood, $h_1$, $h_2$ and $h_3$ the corresponding moisture contents, the exponent n refers to the non-aged wood and the exponent f to the aged wood.

The dimensional changes shall be quantified as a swelling strain ε, difference between the values obtained for two consecutive points of hygroscopic equilibrium, expressed as a percentage in relation to the initial dimension of the interval analysed.

$$\varepsilon_{X,1}^n = \frac{L_2^n - L_1^n}{L_1^n} \quad ; \quad \varepsilon_{X,2}^n = \frac{L_3^n - L_2^n}{L_2^n} \quad \text{for non-aged wood}$$

$$\varepsilon_{X,1}^f = \frac{L_2^f - L_1^f}{L_1^n} \quad ; \quad \varepsilon_{X,2}^f = \frac{L_3^f - L_2^f}{L_2^f} \quad \text{for aged wood:}$$

(2)

where L is the specimen length, X stands for T (tangential direction) or R (radial direction), and the upper scripts have the same meaning as above.

## 3. Results

### 3.1. Equilibrium moisture content

Table 1 gives the equilibrium moisture content values obtained at the three levels of relative humidity studied, for non-aged wood and aged wood. Although the hygroscopicity of radial and tangential specimens is not supposed to be very different, slight differences could be expected due to internal stresses resulting from the heterogeneous swelling behaviour of earlywood and latewood, an effect depending on the specimen orientation. As is made evident in Fig. 6, the hygroscopic behaviour of both types of specimen cannot be considered as significantly different, both for non aged and aged wood.

If we consider the behaviour of non-aged wood as opposed to that of aged wood, it can be seen that in the latter there has been a considerable loss of response with the changes of relative humidity when high values are reached (90.2%), but that this loss of response is lower for lower values of relative humidity (84.2 and 57.6%). However, this fact does not affect the presence of a lineal tendency for the three equilibrium points (Fig. 7). In terms of the species, the values obtained in the hardwoods studied are more homogeneous as a group than those of the conifers, in which *Pinus pinaster* presents values that are quite lower than the values of other pines, although this behaviour changes when aged wood is involved. This fact may be due to an excessive content of impregnation substances with a base of terpenoids and polyphenols.

A comparison of the sorption isotherm slope, calculated according to equation (1), for the two directions once again shows that the behaviour of wood is the same in both cases (Fig. 8). In addition, as a result of the loss of response that occurs in the aged wood for high values of relative humidity, a very slight difference of slope between both RH steps is observed, while in the case of the non-aged wood it is quite marked. In the following, we will not distinguish radial and tangential measurements for the hygroscopic data. The values of sorption isotherm slopes are detailed in Table 2. The values obtained are very homogeneous for all the species with the exception of *Pinus Pinaster* which once again shows anomalous behaviour in relation to the other woods.

Fig. 9 compares the slope of sorption isotherm for aged wood and non-aged wood, calculated according to equation (1), reiterating the fact that the wood is more affected by ageing in the ranges of high humidity than in the ranges of low humidity, where the values of all the species show a very high degree of homogeneity. However, in the ranges of high humidity there is a greater spread of the values obtained, with the differentiation of *Pinus Pinaster* once again standing out.

**3.2 Swelling strain**

Tables 3 and 4 give the swelling strains in radial and tangential directions, respectively, calculated using equation (2), for both non aged and aged wood. In all cases, the swelling of aged wood induced by a given RH step is much lower than that of non aged wood. This is made evident in Figure 10, where all plots lie under the diagonal.

As shown by Fig. 11, swelling anisotropy usually observed in non aged wood, is maintained in aged wood: the tangential swelling is approximately twice the radial swelling, for all the species studied and for both aged and non aged wood. As can be expected, the loss of response in aged wood in terms of moisture content of the wood translates into a loss of dimensional response which once again is more accentuated in the high values of relative air humidity.

In Fig. 12 comparing the swelling strain in both RH intervals, this effect is demonstrated by having the plots corresponding to aged wood close to the diagonal, whereas in non aged wood, they lie well above the diagonal.

## 4. Discussion and conclusions

### 4.1 Definition of ageing coefficients

The results obtained suggest to quantify both types of ageing by appropriate coefficients. The *coefficient of hygroscopic ageing* $\alpha$ will be defined as ratio of sorption isotherm slopes:

$$\alpha_i = \frac{a_i^n}{a_i^f} \tag{3}$$

where i=1,2 for corresponding RH step. According to this definition, the existence of hygroscopic ageing is expressed by a value of this coefficient different from 1. Values higher or lower than one will signify positive or negative hygroscopic ageing, respectively. The *coefficient of dimensional ageing* will be defined as a difference of swelling strains:

$$\eta_{X,i} = \varepsilon_i^n - \varepsilon_i^f \qquad (4)$$

where i=1,2 indicates the RH step and X=R, T the orientation. In this case, positive or negative values of $\eta_X$ signify positive or negative ageing, respectively.

In the right part of Table 2, 3 and 4, the $\alpha$ or $\eta_X$ values have been given, for both RH intervals. In all cases, we have $\alpha > 1$ and $\eta_X > 0$, indicating a positive hygroscopic or dimensional ageing.

**4.2 Hygroscopic ageing**

Fig. 13 shows that hygroscopic ageing is much more pronounced at higher humidity. Moreover, it is stronger in hardwoods than in softwoods. It is possible that this sorptional behaviour of the hardwood woods is in response to a greater neutralisation of the OH groups in their cell wall as a consequence of the larger amount of hemicellulose in their composition. The existence of cellulose and hemicellulose implies the presence of **H-C-OH** and **H-C-CH$_2$OH** groups in the cell wall. These groups have a polar nature and as water is a polar compound, it can be retained by such groups, thereby giving to the cellulose, and therefore by extension to the wood, a hygroscopic nature. Due to its hygroscopic nature, a consequence of the presence of **OH** groups in the cell walls of the vegetal cells that constitute xylematic tissue, the wood constantly tends towards a dynamic equilibrium with the surrounding air. However, this hygroscopicity would remain superficial, external or perimetric with respect to the cell, were it not for the presence of interstices within the cell wall which allow water molecules to enter. This internal surface area varies according to the specific weight of the wood itself and oscillates between 0.2 and $2.8 \times 10^6 \, cm^2/cm^3$, which represents a large number of cavities inside the wood [13]; in this evaluation, the surfaces corresponding to elementary fibrils, micro-fibrils and lumen were considered. The polar nature of the actual water molecules which penetrate the cavities or interstices of the wood causes these molecules to be retained by the **OH** groups of the cellulose molecules which make up the cell wall [20]. The molecules are retained by the cell wall when the distance between the two is less than $10^{-8}$ cm. When any of these OH groups become permanently saturated due to proximity with other groups, the wood begins to lose hygroscopic and dimensional response. This loss of response may be due to the progressive decrease in the concentration of the OH groups during the ageing of the wood, rather than to the degree of crystallisation of the cellulose, as although it changes during the first years of growth, it does not change in a significant way with greater ageing [1].

The retention of water molecules by the **OH** groups in the cell wall is not an anarchic or disorderly process, but is governed by a grouping phenomenon which is a consequence of the polar nature of the bodies in contact. Fixation of water in wood occurs in three different phases according to the entry mechanism: molecular sorption, adsorption and capillary condensation. To a large extent it is precisely the phases of sorption and adsorption which cause the phenomena of hygroscopic and dimensional ageing, as in both phases the water fixing mechanism occurs because of the presence of OH groups. Molecular sorption occurs only for reasons of superficial polarity. This sorption is more accentuated in the amorphous zones of the micellar framework, due to the enormous quantity of free **OH** radicals. At the same time, the water molecules start to position themselves in the crystalline zone, and although it is not possible to fix the point at which all the **OH** groups, both from the amorphous and crystalline zones, are saturated, it appears that this critical point contributes to the total moisture content of the wood at the fibre saturation point about 8 % at most. The second sorption stage is comprised of the **adsorption**. This stage occurs as a consequence of the large number of interstices wood has, or, in other words, the large amount of internal surface area it possesses. This surface area has the form of an intricate submicroscopic system which gives rise to a capillary system capable of retaining the water molecules sorbed in molecular form, which are capable of linking to each other by means of hydrogen bonds. This stage is exothermic, since it has been shown to diminish with increasing temperature. The maximum contribution of this stage to the fibre saturation point is situated at around 12-16 %.

**4.1 Dimensional ageing**

Fig. 14 compares the dimensional ageing for the two intervals of relative humidity and the two directions. Like for hygroscopic ageing, it is more pronounced at high humidity. The dimensional ageing is greater the closer we are to the fibre saturation point. This is due to the loss of response in the cell wall, as many of the OH groups become permanently neutralised, partially recovering their polarity in a small percentage. Although the phenomenon of dimensional ageing must be considered as an ageing process of wood which responds only to a surface area phenomenon, both inter- and intra- microfibrilar in the cell wall, the incidence will vary in function of the direction of observation. For this aspect, the comparison of hardwoods and softwoods is not so clear: in the radial direction hardwoods exhibit generally a higher ageing but in the tangential direction, the dimensional ageing of P. sylvestris et P. pinaster, in the lower range, is higher than all other species.

**4.3. Relation between hygroscopic and dimensional ageing**

Obviously the loss of hygroscopicity, characterised by the hygroscopic ageing, should be responsible for at least a part of the loss of dimensional response, characterised by the dimensional ageing. Fig. 15 evidences a rather good correlation between the hygroscopic and dimensional ageing, for both radial and tangential directions.

For practical purpose, it can be useful to express a dimensional change as a function of the variation of moisture content. This is usually done by calculating swelling coefficients such as:

$$\beta_{X,i}^n = \frac{\varepsilon_i^n}{h_{i+1}^n - h_i^n} \quad ; \quad \beta_{X,i}^f = \frac{\varepsilon_i^f}{h_{i+1}^f - h_i^f} \tag{5}$$

where the subscripts i=1,2 indicate the RH step and X=R, T the direction, and the upperscript n or f refer to the non aged and aged specimens, respectively. Fig. 16a shows that in most cases, the swelling coefficient of aged wood in not very different from that of aged wood, indicating that the cause of dimensional ageing is mostly the reduction of hygroscopicity. There are, however, some exceptions, especially for the higher RH range, where the swelling coefficient of aged wood is much higher or much smaller, resulting in a reduced or triggered dimensional ageing, respectively. Fig. 16b suggests that the linearity between strain and moisture content in the hygroscopic range, usually observed in non-aged wood, is no longer present with aged wood. If it was, the same coefficient should have been measured in both RH ranges. However, in most cases it is higher in the wet than in the dry range, and considerably higher in some cases. This unusual behaviour of the aged wood suggests further investigation with a full record of its swelling and shrinkage behaviour.

Table 1: Equilibrium moisture content of non-aged and aged wood at the three levels of relative humidity

| Moisture content in % | | | | non-aged wood | | | aged wood | | |
|---|---|---|---|---|---|---|---|---|---|
| Species | orientation | | | $H_1$ | $H_2$ | $H_3$ | $H_1$ | $H_2$ | $H_3$ |
| **Pinus sylvestris** | Ps | TG | $\bar{x}$ | 14.05 | 19.19 | 24.67 | 12.78 | 16.18 | 17.74 |
| | | | $\sigma$ | *0.23* | *0.33* | *0.66* | *0.35* | *0.54* | *0.59* |
| | | RD | $\bar{x}$ | 13.64 | 18.48 | 23.63 | 11.61 | 14.44 | 15.79 |
| | | | $\sigma$ | *0.08* | *0.21* | *0.42* | *0.17* | *0.15* | *0.15* |
| **Pinus pinaster** | Pp | TG | $\bar{x}$ | 12.52 | 15.93 | 22.06 | 13.41 | 15.37 | 18.07 |
| | | | $\sigma$ | *0.50* | *0.40* | *0.47* | *0.28* | *0.58* | *0.19* |
| | | RD | $\bar{x}$ | 12.03 | 15.47 | 21.97 | 12.96 | 14.69 | 17.67 |
| | | | $\sigma$ | *0.37* | *0.17* | *0.26* | *0.46* | *0.47* | *0.45* |
| **Pinus insignis** | Pi | TG | $\bar{x}$ | 14.30 | 18.81 | 24.00 | 13.50 | 16.69 | 18.08 |
| | | | $\sigma$ | *0.10* | *0.16* | *0.39* | *0.24* | *0.19* | *0.11* |
| | | RD | $\bar{x}$ | 14.33 | 19.01 | 24.09 | 13.40 | 16.78 | 18.21 |
| | | | $\sigma$ | *0.11* | *0.11* | *0.30* | *0.18* | *0.20* | *0.06* |
| **Populus spp.** | pop | TG | $\bar{x}$ | 12.53 | 17.56 | 23.58 | 12.21 | 15.37 | 16.69 |
| | | | $\sigma$ | *0.08* | *0.26* | *0.23* | *0.29* | *0.28* | *0.30* |
| | | RD | $\bar{x}$ | 12.55 | 17.50 | 23.25 | 12.39 | 15.38 | 16.65 |
| | | | $\sigma$ | *0.08* | *0.18* | *0.30* | *0.25* | *0.19* | *0.22* |
| **Quercus pyrenaica** | Qp | TG | $\bar{x}$ | 13.12 | 17.09 | 22.90 | 12.57 | 14.80 | 16.01 |
| | | | $\sigma$ | *0.10* | *0.12* | *0.29* | *0.10* | *0.08* | *0.13* |
| | | RD | $\bar{x}$ | 13.65 | 17.30 | 22.33 | 12.24 | 14.24 | 15.20 |
| | | | $\sigma$ | *0.08* | *0.20* | *0.28* | *0.12* | *0.18* | *0.24* |
| **Entandrophragma cylindricum** | Ec | TG | $\bar{x}$ | 13.03 | 17.90 | 24.78 | 13.31 | 16.61 | 18.23 |
| | | | $\sigma$ | *0.29* | *0.19* | *0.15* | *0.28* | *0.28* | *0.43* |
| | | RD | $\bar{x}$ | 12.67 | 17.82 | 24.62 | 12.99 | 16.07 | 17.52 |
| | | | $\sigma$ | *0.30* | *0.18* | *0.30* | *0.46* | *0.11* | *0.11* |
| **Chlorophora excelsa** | Ce | TG | $\bar{x}$ | 12.28 | 17.42 | 24.07 | 13.02 | 15.84 | 17.23 |
| | | | $\sigma$ | *0.34* | *0.41* | *0.67* | *0.22* | *0.43* | *0.47* |
| | | RD | $\bar{x}$ | 11.98 | 17.28 | 23.38 | 12.81 | 15.11 | 16.16 |
| | | | $\sigma$ | *0.26* | *0.46* | *0.59* | *0.37* | *0.35* | *0.38* |
| TG = Tangential specimen; RD = Radial specimen<br>Relative Humidity: $H_1$=57.6%; $H_2$=84.2%; $H_3$=90.2% | | | | | | | | | |

Table 2: Slope of sorption isotherm for the two relative humidity steps and corresponding coefficient of hygroscopic ageing

| species | | | slope of sorption isotherm* | | | | Hygroscopic ageing coefficient** | |
|---|---|---|---|---|---|---|---|---|
| | | | non-aged wood | | aged wood | | | |
| | | | relative humidity steps | | | | | |
| | | | $H_1 \rightarrow H_2$ | $H_2 \rightarrow H_3$ | $H_1 \rightarrow H_2$ | $H_2 \rightarrow H_3$ | $H_1 \rightarrow H_2$ | $H_2 \rightarrow H_3$ |
| | | | $a_1^n$ | $a_2^n$ | $a_1^f$ | $a_2^f$ | $\alpha_1$ | $\alpha_2$ |
| SOFTWOOD | Ps | $\bar{x}$ | 0.19 | 0.89 | 0.12 | 0.24 | 1.62 | 3.67 |
| | | $\sigma$ | *0.01* | *0.06* | *0.01* | *0.02* | *0.16* | *0.32* |
| | Pp | $\bar{x}$ | 0.13 | 1.05 | 0.07 | 0.47 | 1.98 | 2.29 |
| | | $\sigma$ | *0.01* | *0.08* | *0.02* | *0.08* | *0.52* | *0.45* |
| | Pi | $\bar{x}$ | 0.17 | 0.86 | 0.12 | 0.23 | 1.40 | 3.70 |
| | | $\sigma$ | *0.00* | *0.05* | *0.01* | *0.03* | *0.10* | *0.50* |
| HARDWOOD | pop | $\bar{x}$ | 0.19 | 0.98 | 0.12 | 0.22 | 1.64 | 4.56 |
| | | $\sigma$ | *0.01* | *0.05* | *0.01* | *0.02* | *0.18* | *0.35* |
| | Qp | $\bar{x}$ | 0.14 | 0.90 | 0.08 | 0.18 | 1.82 | 5.08 |
| | | $\sigma$ | *0.01* | *0.07* | *0.01* | *0.03* | *0.20* | *0.72* |
| | Ec | $\bar{x}$ | 0.19 | 1.14 | 0.12 | 0.26 | 1.60 | 4.51 |
| | | $\sigma$ | *0.01* | *0.03* | *0.02* | *0.03* | *0.24* | *0.55* |
| | Ce | $\bar{x}$ | 0.20 | 1.06 | 0.10 | 0.20 | 2.17 | 5.36 |
| | | $\sigma$ | *0.01* | *0.07* | *0.02* | *0.04* | *0.63* | *0.75* |
| * equation (1); ** equation (3) | | | | | | | | |

Table 3: Tangential swelling strain for the two relative humidity steps and corresponding coefficient of dimensional ageing

| | | | Tangential swelling strain* | | | | Dimensional ageing coefficient** | |
|---|---|---|---|---|---|---|---|---|
| | | | non-aged wood | | aged wood | | | |
| | | | relative humidity steps | | | | | |
| | | | $H_1 \rightarrow H_2$ | $H_2 \rightarrow H_3$ | $H_1 \rightarrow H_2$ | $H_2 \rightarrow H_3$ | $H_1 \rightarrow H_2$ | $H_2 \rightarrow H_3$ |
| species | | | $\varepsilon_{T,1}^n$ | $\varepsilon_{T,2}^n$ | $\varepsilon_{T,1}^f$ | $\varepsilon_{T,2}^f$ | $\eta_{T,1}$ | $\eta_{T,2}$ |
| SOFTWOOD | Ps | $\bar{x}$ | 1.60 | 1.98 | 0.78 | 0.61 | 0.82 | 1.37 |
| | | $\sigma$ | *0.15* | *0.21* | *0.14* | *0.15* | *0.17* | *0.28* |
| | Pp | $\bar{x}$ | 0.88 | 1.95 | 0.15 | 0.98 | 0.73 | 0.98 |
| | | $\sigma$ | *0.17* | *0.13* | *0.07* | *0.18* | *0.15* | *0.19* |
| | Pi | $\bar{x}$ | 1.33 | 1.76 | 1.00 | 0.16 | 0.33 | 1.60 |
| | | $\sigma$ | *0.11* | *0.11* | *0.18* | *0.07* | *0.23* | *0.11* |
| HARDWOOD | pop | $\bar{x}$ | 1.06 | 1.93 | 0.73 | 0.35 | 0.32 | 1.58 |
| | | $\sigma$ | *0.21* | *0.21* | *0.18* | *0.13* | *0.14* | *0.2* |
| | Qp | $\bar{x}$ | 1.60 | 3.04 | 1.38 | 2.01 | 0.22 | 1.04 |
| | | $\sigma$ | *0.20* | *0.20* | *0.23* | *0.24* | *0.18* | *0.15* |
| | Ec | $\bar{x}$ | 2.19 | 3.24 | 1.61 | 0.71 | 0.57 | 2.53 |
| | | $\sigma$ | *0.12* | *0.19* | *0.25* | *0.19* | *0.23* | *0.16* |
| | Ce | $\bar{x}$ | 2.95 | 3.99 | 2.49 | 2.18 | 0.46 | 1.81 |
| | | $\sigma$ | *0.31* | *0.36* | *0.26* | *0.35* | *0.20* | *0.30* |
| * equation (2); ** equation (4) | | | | | | | | |

Table 4: Radial swelling strain for the two relative humidity steps and corresponding coefficient of dimensional ageing

| | | | Radial swelling strain* | | | | Dimensional ageing coefficient | |
|---|---|---|---|---|---|---|---|---|
| | | | non-aged wood | | aged wood | | | |
| | | | relative humidity steps | | | | | |
| | | | $H_1 \rightarrow H_2$ | $H_2 \rightarrow H_3$ | $H_1 \rightarrow H_2$ | $H_2 \rightarrow H_3$ | $H_1 \rightarrow H_2$ | $H_2 \rightarrow H_3$ |
| species | | | $\varepsilon_{R,1}^{n}$ | $\varepsilon_{R,2}^{n}$ | $\varepsilon_{R,1}^{f}$ | $\varepsilon_{R,2}^{f}$ | $\eta_{R,1}$ | $\eta_{R,2}$ |
| S O F T W O O D | Ps | $\bar{x}$ | 0.79 | 0.89 | 0.51 | 0.36 | 0.28 | 0.53 |
| | | $\sigma$ | *0.10* | *0.15* | *0.13* | *0.12* | *0.08* | *0.16* |
| | Pp | $\bar{x}$ | 0.53 | 1.44 | 0.27 | 0.50 | 0.27 | 0.94 |
| | | $\sigma$ | *0.11* | *0.22* | *0.17* | *0.14* | *0.12* | *0.20* |
| | Pi | $\bar{x}$ | 0.55 | 0.66 | 0.39 | 0.29 | 0.17 | 0.37 |
| | | $\sigma$ | *0.08* | *0.08* | *0.09* | *0.07* | *0.05* | *0.09* |
| H A R D W O O D | pop | $\bar{x}$ | 0.72 | 1.59 | 0.44 | 0.36 | 0.29 | 1.23 |
| | | $\sigma$ | *0.15* | *0.33* | *0.10* | *0.16* | *0.16* | *0.41* |
| | Qp | $\bar{x}$ | 0.61 | 1.36 | 0.45 | 0.37 | 0.16 | 1.00 |
| | | $\sigma$ | *0.14* | *0.27* | *0.11* | *0.07* | *0.11* | *0.32* |
| | Ec | $\bar{x}$ | 0.95 | 1.47 | 0.54 | 0.27 | 0.41 | 1.20 |
| | | $\sigma$ | *0.12* | *0.22* | *0.16* | *0.1* | *0.21* | *0.21* |
| | Ce | $\bar{x}$ | 1.20 | 1.40 | 0.95 | 1.05 | 0.24 | 0.35 |
| | | $\sigma$ | *0.21* | *0.17* | *0.15* | *0.09* | *0.15* | *0.15* |
| * equation (2); ** equation (4) | | | | | | | | |

**FIGURE LEGENDS**

Fig. 1:   Isotherm of dynamic hygroscopic states of equilibrium.
Fig. 2:   Sorption and desorption isotherms.
Fig. 3:   Hygroscopic ageing of wood.
Fig. 4:   Characterisation of hygroscopic ageing
Fig. 5:   Preparation of test pieces.
Fig. 6:   Relation between the equilibrium moisture contents of the test pieces with radial and tangential orientation for the 3 levels of relative humidity (57.6%; 84.2% and 90.2% RH): (a) non-aged wood; (b) aged wood
Fig 7:    Relation between equilibrium moisture content for aged wood and non-aged wood, for the three levels of relative humidity
Fig. 8:   Relation between sorption isotherm slope for radial and tangential specimens, for the two steps of relative humidity. Filled symbols: non aged wood; empty symbols: aged wood.
Fig 9:    Relation between slope of sorption isotherm for aged wood and non-aged wood, for the two RH steps.
Fig. 10:  Relationship between the swelling of aged and non aged wood, for the two RH steps and the two orientations.
Fig. 11:  Relationship between the tangential and the radial swelling, for the two RH steps. Filled symbols: non aged wood; empty symbols: aged wood.
Fig.12:   Relationship between the swelling induced by the two RH steps, according to the orientation and the ageing state.
Fig 13:   Relationship between the coefficient of hygroscopic ageing in the two ranges of relative humidity.
Fig 14:   Relationship between the coefficient of dimensional ageing (a) in the two ranges of relative humidity, and (b) for the two directions
Fig 15:   Relation between coefficients of hygroscopic and dimensional ageing
Fig 16:   Effect of the ageing treatment on the swelling coefficient of wood in both radial (Rd) and tangential (Tg) directions: (a) aged versus non aged wood ; (b) wet range versus dry range of relative humidity.

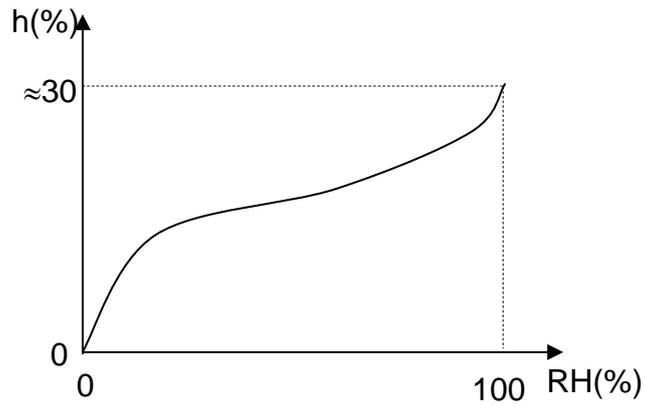

**Fig. 1: Isotherm of dynamic hygroscopic states of equilibrium.**

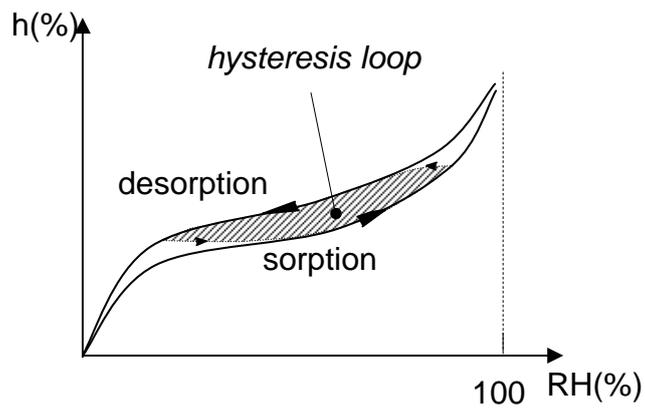

**Fig. 2: Sorption and desorption isotherms.**

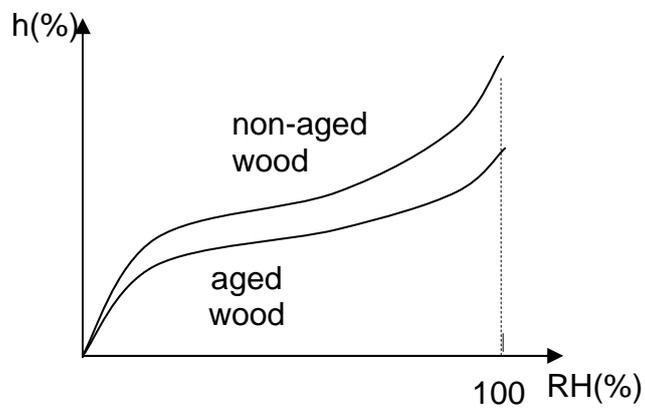

**Fig. 3: Hygroscopic ageing of wood.**

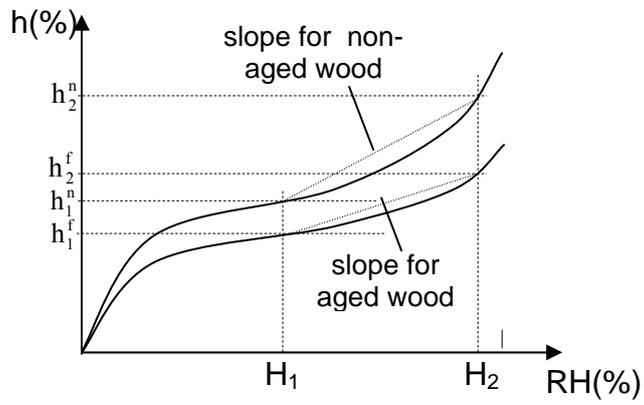

**Fig. 4: Characterisation of hygroscopic fatigue**

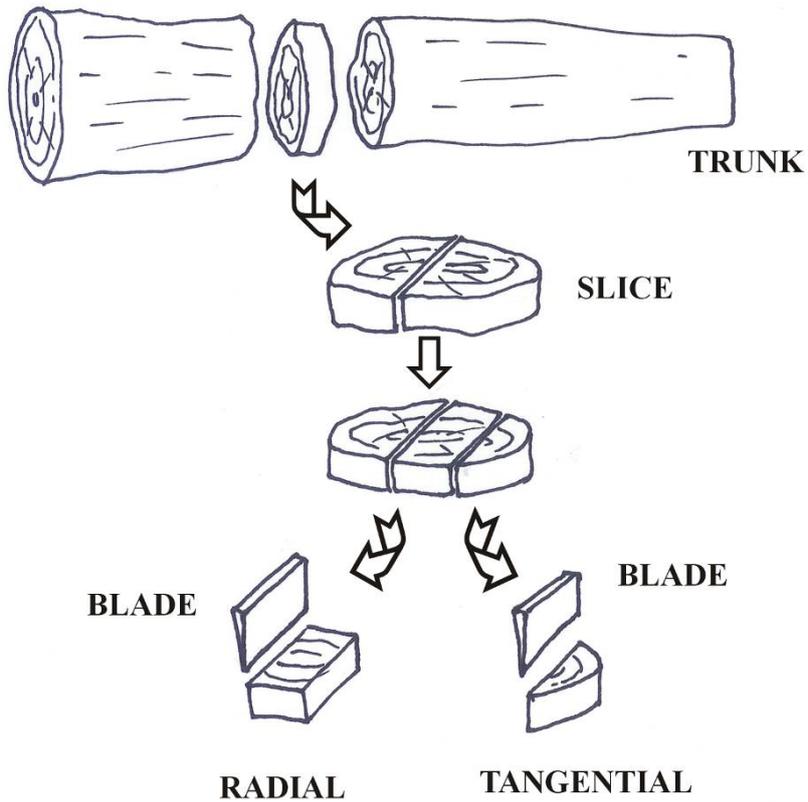

**Fig. 5: Preparation of test pieces.**

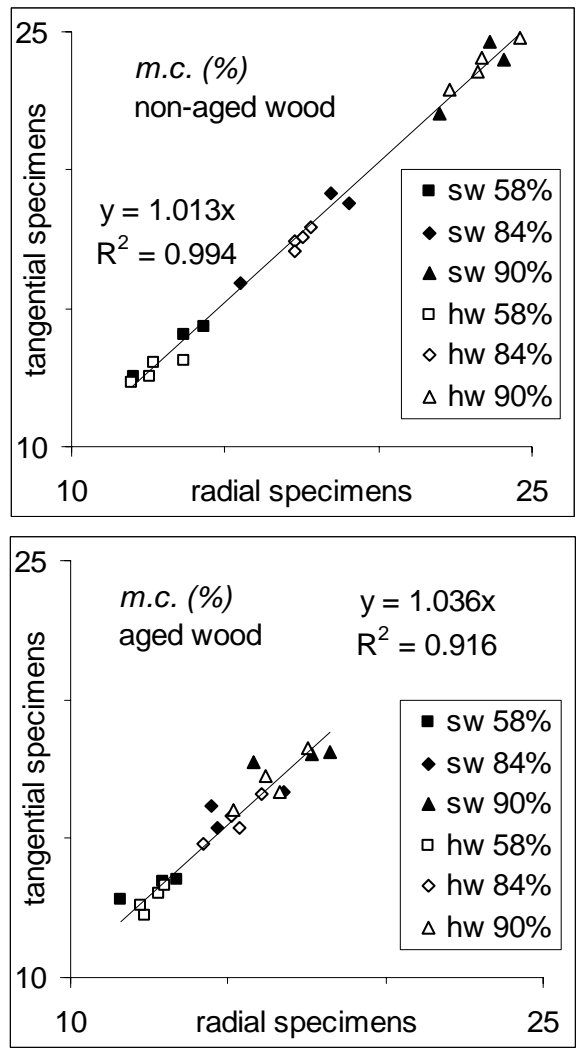

**Fig. 6: Relation between the equilibrium moisture contents of the test pieces with radial and tangential orientation for the 3 levels of relative humidity (57.6%; 84.2% and 90.2% RH): (a) non-fatigued wood; (b) fatigued wood**

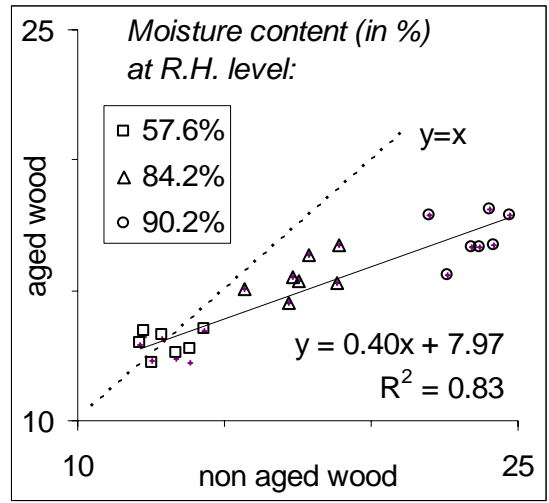

**Fig 7: Relation between equilibrium moisture content for fatigued wood and non-fatigued wood, for the three levels of relative humidity**

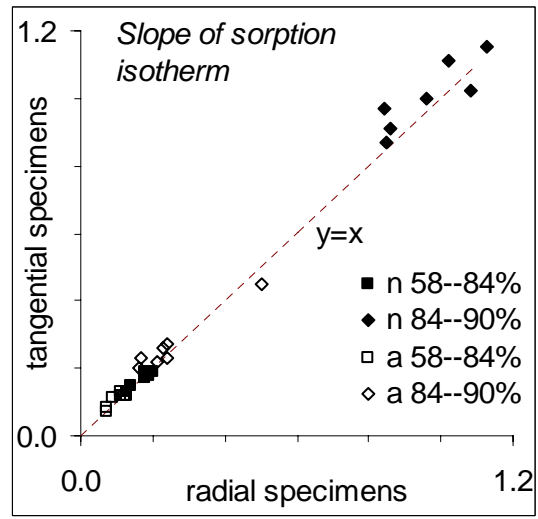

**Fig. 8: Relation between sorption isotherm slope for radial and tangential specimens, for the two steps of relative humidity. Filled symbols: non fatigued wood; empty symbols: fatigued wood.**

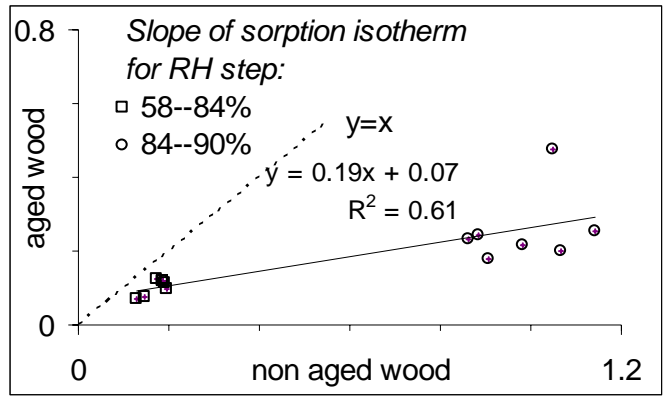

**Fig 9: Relation between slope of sorption isotherm for fatigued wood and non-fatigued wood, for the two RH steps.**

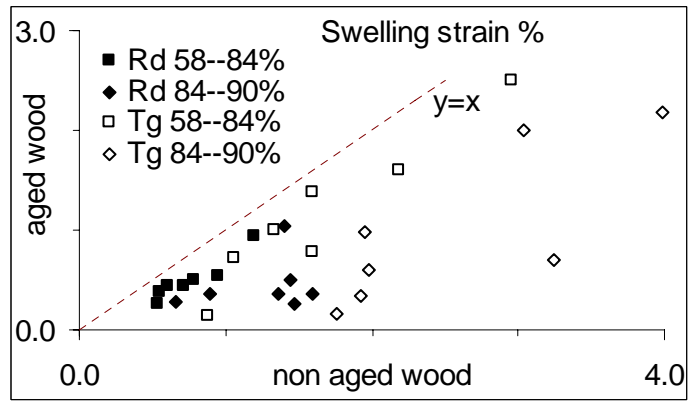

**Fig. 10: Relationship between the swelling of fatigued and non fatigued wood, for the two RH steps and the two orientations.**

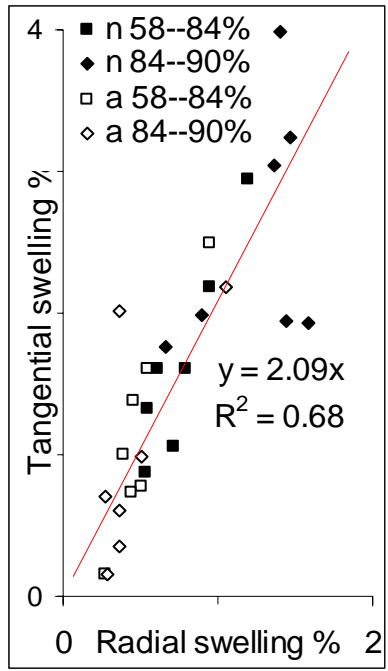

**Fig 11: Relationship between the tangential and the radial swelling, for the two RH steps. Filled symbols: non fatigued wood; empty symbols: fatigued wood.**

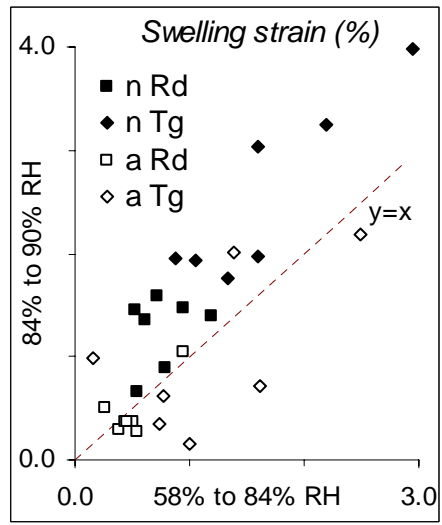

**Fig 12: Relationship between the swelling induced by the two RH steps, according to the orientation and the fatigue state.**

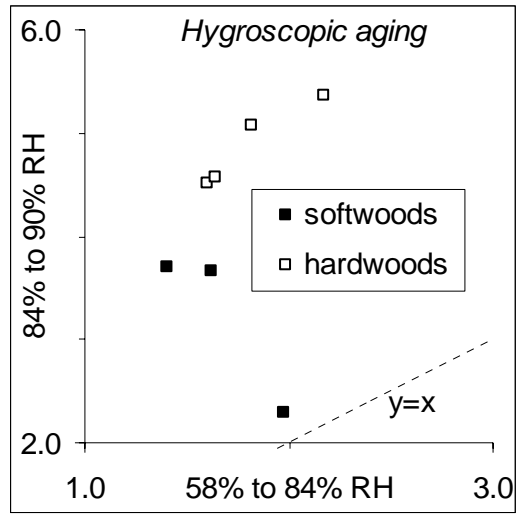

**Fig 13: Relationship between the coefficient of hygroscopic fatigue in the two ranges of relative humidity.**

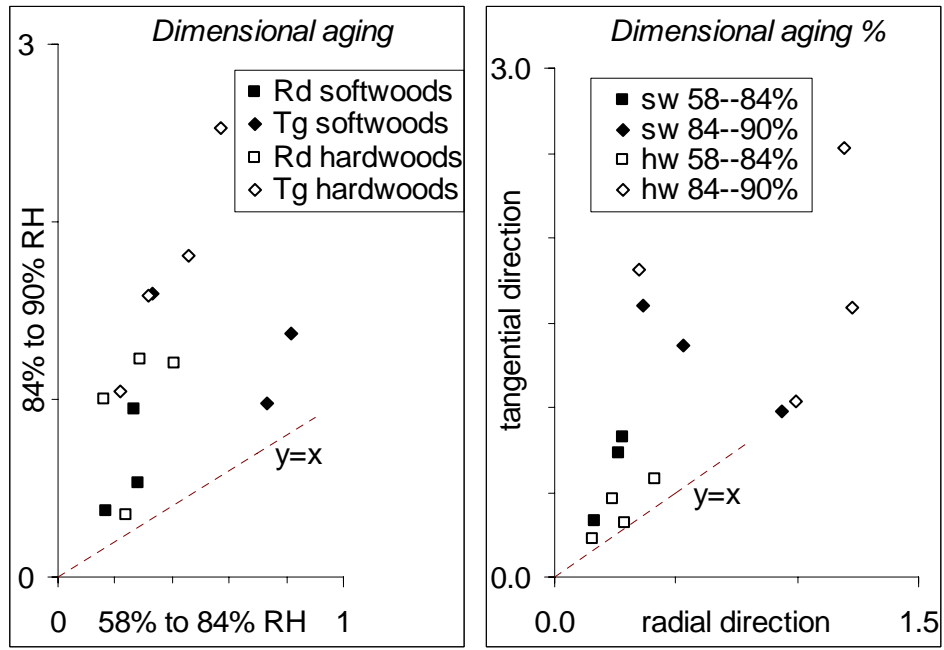

**Fig 14: Relationship between the coefficient of dimensional fatigue (a) in the two ranges of relative humidity, and (b) for the two directions**

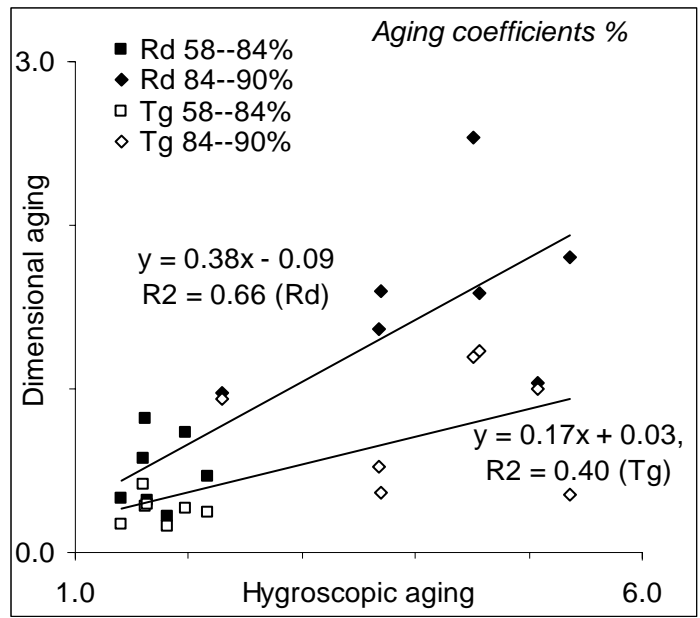

**Fig 15: Relation between coefficients of hygroscopic and dimensional fatigue**

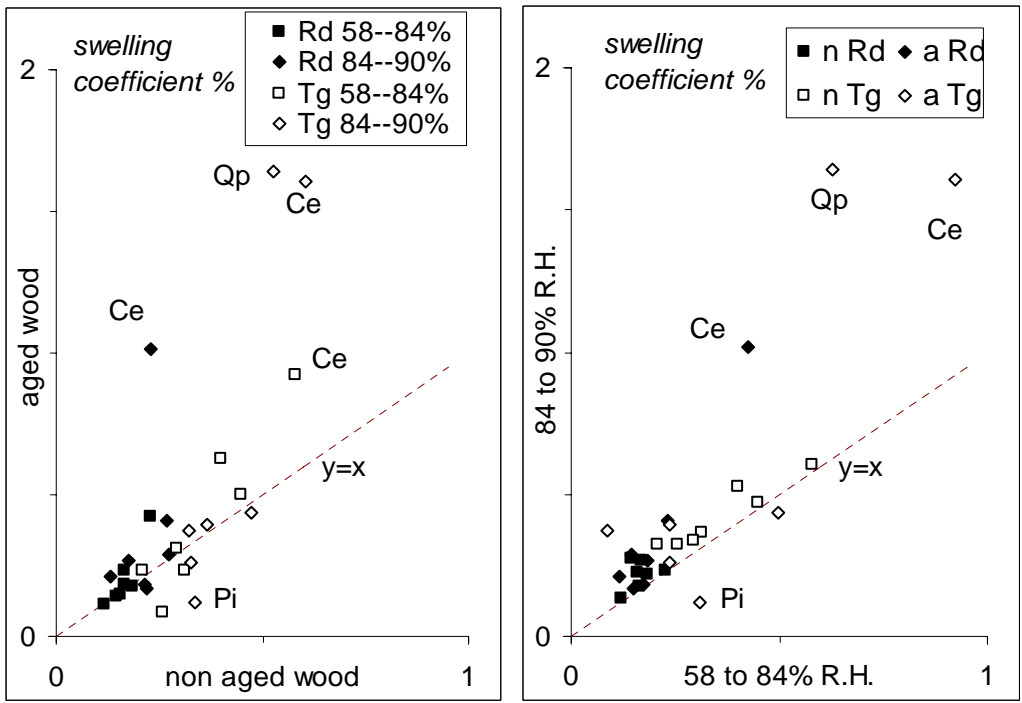

**Fig 16: Effect of the fatigue treatment on the swelling coefficient of wood in both radial (Rd) and tangential (Tg) directions: (a) fatigued versus non fatigued wood ; (b) wet range versus dry range of relative humidity.**